\documentclass[nofootinbib,twocolumn,prd,preprintnumbers,superscriptaddress]{revtex4-1}

\usepackage{amsfonts}
\usepackage{graphicx}
\usepackage[colorlinks=true,linkcolor=blue,citecolor=teal,urlcolor=blue]{hyperref}
\usepackage{color}
\usepackage[dvipsnames]{xcolor}
\usepackage{amsmath}
\usepackage{cases}
\usepackage{braket}
\usepackage[T1]{fontenc}
\usepackage{mathrsfs}
\usepackage{enumerate}
\usepackage{bm}
\usepackage{multirow}
\usepackage{tabularx}
\usepackage{comment}
\usepackage{mathrsfs}
\usepackage[scr=boondoxo]{mathalfa}

\begin{document}

\title{Can the NANOGrav observations constrain the geometry of the universe?}

\author{Matteo Califano}
\email{matteo.califano@unina.it}
\affiliation{Scuola Superiore Meridionale, Largo San Marcellino 10, 80138 Napoli, Italy}
\affiliation{Istituto Nazionale di Fisica Nucleare (INFN), Sezione di Napoli, Via Cinthia 21, 80126 Napoli, Italy}

\author{Rocco D'Agostino}
\email{rocco.dagostino@unina.it}
\affiliation{Scuola Superiore Meridionale, Largo San Marcellino 10, 80138 Napoli, Italy}
\affiliation{Istituto Nazionale di Fisica Nucleare (INFN), Sezione di Napoli, Via Cinthia 21, 80126 Napoli, Italy}

\author{Daniele Vernieri}
\email{daniele.vernieri@unina.it}
\affiliation{Dipartimento di Fisica ``E. Pancini'', Universit\`a di Napoli ``Federico II", Via Cinthia 21, 80126 Napoli, Italy}
\affiliation{Istituto Nazionale di Fisica Nucleare (INFN), Sezione di Napoli, Via Cinthia 21, 80126 Napoli, Italy}
\affiliation{Scuola Superiore Meridionale, Largo San Marcellino 10, 80138 Napoli, Italy} 

\begin{abstract}

The theory of inflation provides an elegant explanation for the nearly flat universe observed today, which represents one of the pillars of the standard cosmological model. However, recent studies have reported some deviations from a flat geometry, arguing that a closed universe would be instead favored by observations. Given its central role played in the cosmological context, this paper revisits the issue of spatial curvature in light of the stochastic gravitational wave background signal recently detected by the NANOGrav collaboration. For this purpose, we investigate the primordial gravitational waves generated during inflation and their propagation in the post-inflationary universe. We propose a new parametrization of the gravitational wave power spectrum, taking into account spatial curvature, the tensor-to-scalar ratio and the spectral index of tensor perturbations. Therefore, we compare the theoretical predictions with NANOGrav data to possibly constrain the geometry of the universe. We find that the choice of the priors has a significant effect on the computed posterior distributions.
In particular, using flat uniform priors results in $\Omega_{\mathcal{K},0}= 0.00 \pm 0.67$ at the 68\% confidence level.  On the other hand, imposing a Planck prior, we obtain $\Omega_{\mathcal{K},0}= -0.05 \pm 0.17$ at the 68\% confidence level. This result aligns with the analysis of the cosmic microwave background radiation, and no deviations from a flat universe are found.

\end{abstract}

\maketitle

\section{Introduction}

The inclusion of the cosmological constant ($\Lambda$) into Einstein's equations of general relativity (GR) offers the simplest way to explain the current acceleration of the universe suggested by observations over the last two decades \cite{Carroll:1991mt, Riess:1998cb, Perlmutter:1998np, Peebles:2002gy}. The cosmic speed-up can be ascribed, more generally, to a mysterious fluid with negative pressure, known as dark energy, which propels the late-time dynamics \cite{Copeland:2006wr, Frieman:2008sn, Weinberg:2013agg}.  Additionally, the total matter content in the universe is constituted of only a small fraction of baryons, while the main part is dominated by hypothetical dark matter particles that do not interact with the electromagnetic field \cite{Bond:1984fp, Persic:1992hci, DAgostino:2022ckg}. 
On the other hand, the formation of cosmic structures arises from primordial quantum fluctuations that were stretched beyond the horizon during inflation \cite{Guth:1980zm}.  Subsequently, these fluctuations re-entered the horizon as density perturbations, giving rise to all the structures present in the universe \cite{Mukhanov:1990me}. The inflationary mechanism is responsible for the nearly vanishing curvature, isotropy and homogeneity on large scales observed today \cite{Linde:1981mu}.
Overall, such a picture of the universe goes under the name of $\Lambda$-Cold Dark Matter $(\Lambda$CDM) scenario, which stands out as the standard model of cosmology \cite{WMAP:2012nax, Planck:2018vyg}. 

Whilst the $\Lambda$CDM model has proven successful in explaining most observations theoretical limitations regarding the nature of the dark components cast doubt on its confirmation as the ultimate picture of the cosmos. 
Indeed, the debated origin of $\Lambda$, interpreted as the vacuum energy density, leads to the well-known fine-tuning problem \cite{Weinberg:1988cp, Padmanabhan:2002ji, DAgostino:2022fcx}. 
At the same time, recent tensions among cosmic data have questioned the validity of the standard paradigm to thoroughly describe the evolution of the universe \cite{DiValentino:2020vvd, Riess:2021jrx, Perivolaropoulos:2021jda, DAgostino:2023cgx}.
All this has motivated through the years the search for possible alternatives. Several examples, among the others, include dynamically evolving scalar fields \cite{Ratra:1987rm, Caldwell:1997ii, Tsujikawa:2013fta}, unified dark models \cite{Kamenshchik:2001cp, Scherrer:2004au, Capozziello:2017buj, Capozziello:2018mds, DAgostino:2021vvv}, or holographic dark energy \cite{Li:2004rb, DAgostino:2019wko}. Alternatively, it is possible to explain the universe's acceleration using higher-order curvature invariants \cite{Sotiriou:2008rp, DeFelice:2010aj, Clifton:2011jh, Nojiri:2017ncd, DAgostino:2019hvh, Capozziello:2019cav, Bajardi:2022tzn}, non-local modifications of gravity \cite{Deser:2007jk, Maggiore:2014sia, Capozziello:2022rac, Capozziello:2023ccw}, or even by considering different geometric structures of spacetime, e.g., based on torsion \cite{Bengochea:2008gz, Linder:2010py, DAgostino:2018ngy, CANTATA:2021ktz} or non-metricity \cite{BeltranJimenez:2019tme,Atayde:2021pgb, Capozziello:2022wgl, DAgostino:2022tdk}. 

Furthermore, the characteristic vanishing curvature of the standard cosmological paradigm has been put into question due to some inconsistencies that recently emerged between the cosmic microwave background (CMB) data and baryon acoustic oscillation (BAO) measurements \cite{Park:2017xbl, Handley:2019tkm, ODwyer:2019rvi}. These inconsistencies might be traced back to the assumption of a flat $\Lambda$CDM model as the fiducial cosmology. In fact, when the flatness condition is relaxed, the combined analysis of BAO and CMB data would indicate evidence for a closed universe at the $2\sigma$ confidence level (CL) \cite{Glanville:2022xes}. 

The presence of non-zero curvature manifests its influence not only at the background but also at linear perturbations, inducing modifications to transfer functions and power spectra of scalar and tensor perturbations generated in the inflationary era \cite{Lewis:1999bs, DAgostino:2023tgm}. Moreover, deviations from a flat geometry would change the duration of inflation itself, potentially offering a consistent description of the CMB large-scale amplitudes \cite{Efstathiou:2003hk, Lasenby:2003ur}.
For these reasons, constraining the geometry of the universe becomes a fundamental task of modern cosmology. 

To address this issue, in the present study, we analyze the effects of spatial curvature on the stochastic gravitational wave background (GWB). 
The latter represents a crucial prediction for the theory of inflation and allows for probing energy scales beyond the standard achievable in experiments of particle physics. 
Specifically, we focus on the recent detection of a nHz GWB signal by the North American Nanohertz Observatory for Gravitational Waves (NANOGrav), based on the 15-year Pulsar Timing Array (PTA) data \cite{NANOGrav:2023gor}. Albeit the measured GWB amplitude and spectrum are compatible with the astrophysical signal from a supermassive black-hole binary (SMBHB) population, nevertheless alternative astrophysical or cosmological sources cannot be fully discarded. 
Ref.~\cite{NANOGrav:2023hvm} showed that several cosmological models are actually able to reproduce the observed GWB signal. In particular, the latter could be suitably interpreted within the framework of inflation, domain walls, scalar-induced GWs and first-order phase transition scenarios. These seem to be statistically favored with respect to the standard SMBHB interpretation, although conclusive evidence for new physics is still premature.

This paper is organized as follows. In Sec.~\ref{sec:theory}, we overview the solutions of the Friedmann equations in different cosmic eras, for different spatial geometries, and we analyze tensor perturbations describing the propagation of primordial GWs. In Sec.~\ref{sec:GW power spectrum}, we discuss the evolution of transfer functions in the post-inflationary universe, and we derive the power spectrum and energy density of primordial GWs. In Sec.~\ref{sec:methodology}, we describe the methodology we employ to analyze the NANOGrav data. In Sec.~\ref{sec:results}, we present the constraints on the curvature density parameter and discuss our results in light of previous findings in the literature. 
In Sec.~\ref{sec:conclusions}, we conclude with the summary of our main findings and the final remarks.

In this work, we use units such that $c=\hbar=8\pi G=1$, and the metric signature $(-,\,+,\,+,\,+)$.

\section{Theoretical setup}
\label{sec:theory}

We start by considering the Einstein field equations 
\begin{equation}
    R_{\mu\nu}-\dfrac{R}{2}g_{\mu\nu}+\Lambda g_{\mu\nu}= T_{\mu\nu}\,,
    \label{eq:EFE}
\end{equation}
where $R_{\mu\nu}$ and $R$ are the Ricci tensor and scalar, respectively, $g_{\mu\nu}$ is the metric tensor, and $T_{\mu\nu}$ is the energy-momentum tensor of matter fields. 

According to the cosmological principle, the background dynamics of a homogeneous and isotropic Universe can be described by using the Friedmann-Lema\^itre-Robertson-Walker (FLRW) metric:
\begin{equation}
     ds^2=a(\tau)^2\left(-d\tau^2+\gamma_{ij}dx^idx^j\right),
     \label{eq:FRW metric}
\end{equation}
where $a(\tau)$ is the scale factor as a function of the conformal time, $\tau$. Here, $\gamma_{ij}$ is the metric of the spatial hypersurface:
\begin{equation}
   \gamma_{ij}dx^idx^j=\dfrac{dr^2}{1-\mathcal{K} r^2}+r^2\left(d\theta^2+\sin^2\theta\, d\phi^2\right),
\end{equation}
where $\mathcal{K}$ is the curvature parameter\footnote{Notice that, in our notation, $\mathcal{K}$ has units of \textbf{length}$^{-2}$.} that describes the geometry of the 3D space, with $\mathcal{K}=0$ corresponding to a flat (Euclidean) universe, while $\mathcal{K}>0$ and $\mathcal{K}<0$ to closed (spherical) and open (hyperbolic) universes, respectively. 
If we assume that the matter content of the universe is in the form of a perfect fluid of density $\rho$ and pressure $p$, we can write 
\begin{equation} 
    T_{\mu\nu}= \rho\left(1+w\right)u_{\mu}u_{\nu}+p g_{\mu\nu}\,,
\end{equation}
where $w\equiv p/\rho$ is the barotropic equation of state parameter, and $u_{\mu}$ is the fluid four-velocity. 
Under the given assumptions, we obtain the Friedmann equations
\begin{align}
     \mathcal{H}^2&=\dfrac{1}{3}\left(\rho +\Lambda\right)a^2-\mathcal{K}\,, \label{eq: friedmann 1} \\
   \mathcal{H}'+\mathcal{H}^2&=\dfrac{1}{6}\left(\rho-3p\right)a^2+\dfrac{2 \Lambda }{3}a^2-\mathcal{K}\,,
   \label{eq: friedmann 2}
\end{align}
where $\mathcal{H}\equiv a'/a$ is the conformal Hubble parameter, with the prime denoting the derivative with respect to $\tau$. Additionally, the conservation of the energy-momentum tensor results in the continuity equation
\begin{equation}
    \rho'+3\mathcal{H}(1+w)\rho=0\,.
    \label{eq:continuity}
\end{equation}
The latter can be combined with Eqs.~\eqref{eq: friedmann 1} and \eqref{eq: friedmann 2} to obtain the solutions of the scale factor in a given cosmological model.

In particular, a de Sitter inflationary universe, for which $w=-1$, admits the following solution \cite{DAgostino:2023tgm}:
\begin{numcases} {a_{\inf}(\tau)=}
      -\dfrac{\sqrt{|\mathcal{K}|}}{ \mathcal H_{\Lambda}\sinh\left(\tau\sqrt{|\mathcal{K}|}\right)}\,, \hspace{0.3cm} &$\mathcal{K}<0$\,,\label{eq: openINF}\\  -\dfrac{1}{\mathcal H_{\Lambda}\tau }\,, \hspace{0.3cm} &$\mathcal{K}=0$\,,\label{eq: flatINF}\\
      -\dfrac{\sqrt{\mathcal{K}}}{\mathcal H_{\Lambda}\sin\left(\tau\sqrt{\mathcal{K}}\right)}\,,\hspace{0.3cm} &$\mathcal{K}>0$\,. \label{eq: closedINF}
\end{numcases}
where $\mathcal{H}_\Lambda\equiv\sqrt{\Lambda/3}$.
In the radiation-dominated (RD) epoch, $w=1/3$ and $\Lambda=0$, we get
\begin{numcases}{a_\text{RD}(\tau)\propto}
  \sinh\left(\tau\sqrt{|\mathcal{K}|}\right)\,, \hspace{0.3cm} &$\mathcal{K}<0$\,,\label{eq: radOpen}\\ 
       \tau\,, \hspace{0.3cm} &$\mathcal{K}=0$\,,\label{eq: radFlat}\\
\sin\left(\tau\sqrt{\mathcal{K}}\right)\,, \hspace{0.3cm} &$\mathcal{K}>0$\,. \label{eq: radClosed}
\end{numcases}
Additionally, in the matter-dominated (MD) epoch, $w=0$ and $\Lambda=0$, one finds 
 \begin{numcases}{ a_\text{MD}(\tau)\propto}
\cosh \left(\tau\sqrt{|\mathcal{K}|}\right)-1, \hspace{0.3cm} & $\mathcal{K}<0$\,,\label{eq: matOpen}\\
       \tau^2\,, \hspace{0.3cm} & $\mathcal{K}=0$\,,\label{eq: matFlat}\\
    1-\cos\left(\tau\sqrt{\mathcal{K}}\right), \hspace{0.3cm} & $\mathcal{K}>0\,.$\label{eq: matClosed}
\end{numcases}

\subsection{Tensor perturbations}

To study the primordial power spectrum of GWs, we consider linear perturbations around the FLRW metric:
\begin{equation}
    ds^2= a(\tau)^2\left[-d\tau^2+(\gamma_{ij}+h_{ij})dx^idx^j\right].
    \label{perturbed metric}
\end{equation}
Here, $h_{ij}$ are small tensor perturbations satisfying $h_i^i=\mathscr{D}_ih_j^i=0$, where $\mathscr{D}_i$ indicates the $\gamma_{ij}$-compatible covariant derivative.
Within this framework, the GW evolution is governed by \cite{Mukhanov:1990me}
\begin{equation}\label{GW evolution}
    h^{\prime\prime}_{ij}+2\mathcal{H}h^{\prime}_{ij}+2\mathcal{K}h_{ij}=\mathscr{D}^2 h_{ij}\,,
\end{equation}
where $\mathscr{D}^2\equiv\gamma^{ij}\mathscr{D}_i\mathscr{D}_j$.
The above equation could be solved through the expansion
\begin{equation}
     h_{ij}(\tau)=\sum_s\sum_{nlm}h_{\mathscr{k}\ell m}^{(s)}(\tau)\, Q_{ij}^{\mathscr{k}\ell m(s)}(r,\theta,\phi) \,,
     \label{eq:tensor decomposition}
\end{equation}
with $Q^{\mathscr{k}\ell m(s)}_{ij}$ being tensor harmonics defined as
\begin{equation}
    \mathscr{D}^2Q^{\mathscr{k}\ell m(s)}_{ij}=-(\mathscr{k}^2-3\mathcal{K})Q^{\mathscr{k}lm(s)}_{ij}\,.
    \label{eq: laplacian}
\end{equation}
Here, we have introduced the curved-space wavenumber, $\mathscr{k}=\sqrt{k^2+3\mathcal{K}}$, which reduces to the flat Fourier eigenmode, $k$, in the limit $\mathcal{K}\rightarrow 0$. 
The completeness of the tensor harmonic spectrum requires $\mathscr{k}/\sqrt{\mathcal{K}}=3,4,5,\hdots$\,, $2\leq \ell \leq \mathscr{k}-1$ and $-\ell\leq m \leq \ell$, for $\mathcal{K}>0$, while $\mathscr{k}\geq 0$, $\ell\geq 2$ and $-\ell\leq m \leq \ell$, for $\mathcal{K}\leq 0$ . Furthermore, $s$ refers to the parity of harmonics (see Refs.~\cite{Abbott:1986ct,Hu:1997mn,Akama:2018cqv} for details).

Therefore, the GW evolution in a spatially curved universe is obtained by solving the master equation \cite{DAgostino:2023tgm}
\begin{equation}
\sigma_\mathscr{k}^{\prime\prime}+\left(\mathscr{k}^2-\mathcal{K}-\dfrac{a''}{a}\right)\sigma_\mathscr{k}=0\,,
\label{eq:master}
\end{equation}
where the eigenmodes $\sigma_\mathscr{k}(\tau)\equiv a(\tau) h_\mathscr{k}(\tau)$ are subjected to the normalization  
\begin{equation}
    \sigma_\mathscr{k} \sigma_\mathscr{k}^{*\prime}-\sigma_\mathscr{k}^{\prime}\sigma_\mathscr{k}^{*}=i\,.
    \label{eq:normalization}
\end{equation}
Notice that we have dropped the indexes $\{n,l,m,s\}$ for the sake of brevity.
As shown in Ref.~\cite{DAgostino:2023tgm}, the solutions of Eq.~\eqref{eq:master} are
\begin{numcases}{\sigma_\mathscr{k}(\tau)=e^{-i \mathscr{k}\tau}}
    \dfrac{\mathscr{k}-i\sqrt{|\mathcal{K}|} \coth\left(\tau\sqrt{|\mathcal{K}|}\right)}{\sqrt{2\mathscr{k} \left(\mathscr{k}^2+|\mathcal{K}|\right)}}\,, \ \mathcal{K}<0\,, \\
    \frac{1}{\sqrt{2\mathscr{k}}}\left(1-\dfrac{i}{\mathscr{k} \tau}\right), \hspace{1.55cm} \mathcal{\mathcal{K}}=0\,, \\
     \dfrac{\mathscr{k}-i\sqrt{\mathcal{K}} \cot\left(\tau\sqrt{\mathcal{K}}\right) }{\sqrt{2\mathscr{k}(\mathscr{k}^2-\mathcal{K})}}\,, \hspace{0.8cm}\mathcal{\mathcal{K}}>0\,.
\end{numcases}
It is straightforward to verify that the open and closed cases reduce to the flat solution in the limit for $\mathcal{K}\rightarrow 0$.

\section{Primordial Gravitational Waves}
\label{sec:GW power spectrum}

The primordial power spectrum is provided by tensor perturbations generated during the inflationary epoch. In general, one may write the solution of tensor perturbations, at a given time, as \cite{Watanabe:2006qe,Saikawa:2018rcs,Bernal:2019lpc}
\begin{equation}
    h_\mathscr{k}(\tau)\equiv h_{\mathscr{k},\text{inf}}(\tau)\,\mathscr{T}_\mathscr{k}(\tau)\,,
    \label{eq: factorized solution}
\end{equation}
where $h_{\mathscr{k},\text{inf}}$ is the amplitude of GWs that left the horizon during inflation, while $\mathscr{T}_\mathscr{k}$ is the transfer function describing the evolution of GWs after inflation, such that $\mathscr{T}_\mathscr{k}\rightarrow 1$ for $\mathscr{k}\ll \mathcal{H}$. 
Specifically, the transfer function is obtained by solving the equation
\begin{equation}
    \mathscr{T}_\mathscr{k}''(\tau)+2\mathcal{H}\mathscr{T}_\mathscr{k}'(\tau)+\left( \mathscr{k}^2-\mathcal{K}\right)\mathscr{T}_\mathscr{k}(\tau)=0\,, \label{eq: curved GW eq TF}
\end{equation}
together with the boundary conditions  $\mathscr{T}_\mathscr{k}(0)=1$ and $\mathscr{T}_\mathscr{k}'(0)=0$. 
Eq.~\eqref{eq: curved GW eq TF} describes the radiation\,(matter) epoch for $\tau<\tau_\text{eq}\, (\tau>\tau_\text{eq})$ or, equivalently, $\mathscr{k}<\mathscr{k}_\text{eq}\,(\mathscr{k}>\mathscr{k}_\text{eq})$, where $\tau_\text{eq}$ is the matter-radiation equivalence time. 

The full derivation of the transfer functions in the different cosmological epochs, for different spatial geometries, is given in Ref.~\cite{DAgostino:2023tgm}. Specifically, one can show that the transfer function in the RD epoch is given by
\begin{numcases}{\mathscr{T}_{\text{RD}}=}
 \frac{\sqrt{| \mathcal K | } \sin (\mathscr{k} \tau)}{\mathscr{k} \sinh\left(\tau \sqrt{| \mathcal K | }\right)}\,, \quad \mathcal{K}<0\,, \label{eq:TF_RD_open}\\
      \dfrac{\sin(\mathscr{k}\tau)}{\mathscr{k}\tau}\,, \hspace{1.65cm} \mathcal{K}=0\,, \label{eq:TF_RD_flat}\\
     \frac{\sqrt{\mathcal K } \sin (\mathscr{k} \tau)}{\mathscr{k} \sin \left(\tau \sqrt{\mathcal K } \right)}\,, \hspace{0.85cm} \mathcal{K}>0\,.
\label{eq:TF_RD_closed}
\end{numcases}  
Moreover, in the MD epoch, we find
\begin{numcases}{\mathscr{T}_{\text{MD}}=}
 \frac{3 | \mathcal K | ^{3/2} \sin (\mathscr k \tau) \coth \left(\frac{\tau \sqrt{| \mathcal K  | }}{2}\right)-6 \mathscr k | \mathcal K  |  \cos (\mathscr k \tau)}{\mathscr{k} \left(4 \mathscr k^2+|\mathcal K |\right) \left[\cosh \left(\tau \sqrt{| \mathcal K  | }\right)-1\right]}\,, \label{eq:TF_MD_open}\\
      \frac{3}{(\mathscr{k}\tau)^3}\left[\sin(\mathscr{k}\tau)-\mathscr{k}\tau \cos(\mathscr{k}\tau)\right]\,, \label{eq:TF_MD_flat}\\
     \frac{6 \mathscr{k} \mathcal K    \cos (\mathscr{k} \tau)-3 \mathcal K  ^{3/2} \sin (\mathscr k \tau) \cot \left(\frac{ \tau \sqrt{\mathcal K } }{2}\right)}{\mathscr k \left(4 \mathscr k^2-\mathcal K  \right) \left[\cos \left( \tau \sqrt{\mathcal K  }\right)-1\right]}\,, 
\label{eq:TF_MD_closed}
\end{numcases}  
for an open, flat and closed universe, respectively. The matching between the RD and MD epochs is obtained by considering radiation modes smoothly propagating into the matter era (see Ref.~\cite{DAgostino:2023tgm} for the details).

Let us now consider the energy density of GWs \cite{Watanabe:2006qe}:
\begin{equation}
    \rho_\text{GW}(\tau)=\dfrac{\langle h_{ij}'(\tau)h^{ij\prime}(\tau)\rangle}{4a(\tau)^2}\,,
\end{equation}
where $\langle\hdots\rangle$ indicates the spatial average over different wavelengths.  Assuming primordial GWs to be unpolarized and using Eq.~\eqref{eq: factorized solution}, one finds
\begin{equation}
    \rho_\text{GW}(\tau)=\dfrac{1}{2a(\tau)^2}\int d\ln \mathscr{k}\, \mathcal{P}_T(\mathscr{k})\,|\mathscr{T}'_\mathscr{k}(\tau)|^2\,,
\end{equation}
where the primordial power spectrum is defined by
\begin{equation}
    \mathcal{P}_T(\mathscr{k})\equiv \dfrac{\mathscr{k}^3}{\pi^2}|h_{\mathscr{k},\text{inf}}|^2\,.
    \label{eq: PPS2}
\end{equation}
Therefore, the GW spectral density can be written as
\begin{equation}
\Omega_\text{GW}(\tau)\equiv \dfrac{1}{\rho_\text{cr}(\tau)}\dfrac{d\rho_\text{GW}(\tau)}{d\ln \mathscr{k}}=
\dfrac{\mathcal{P}_T(\mathscr{k})}{12 \mathcal{H}(\tau)^2}\left|\mathscr{T}_\mathscr{k}^{\prime}(\tau)\right|^2\,,
\label{eq: spectral energy density}
\end{equation}
where $\rho_\text{cr}(\tau)\equiv 3\mathcal{H}(\tau)^2 a(\tau)^{-2}$ is the critical density of the universe.
In particular, for a spatially curved de Sitter universe, we find \cite{DAgostino:2023tgm}
\begin{equation}
    \mathcal P_T(\mathscr{k})= \mathcal{P}_{T,\text{flat}}\, \frac{\mathscr{k}^4}{\mathscr{k}^4-\mathcal{K}^2}\,,
    \label{eq:curved PS}
\end{equation}
where $\mathcal P_{T,\text{flat}}=\left(\mathcal{H}_\Lambda/\pi\right)^2$ is the primordial power spectrum  for a flat geometry. This
can be parametrized by adopting the typical power-law form \cite{Planck:2013pxb}
\begin{equation}
   \mathcal{P}_{T,\text{flat}}(\mathscr k)= r A_S \left(\frac{\mathscr{k}}{\mathscr{k}_0}\right)^{n_T}\,,
   \label{parametrization flat PS}
\end{equation}
where $r\equiv A_T/A_S$ is the tensor-to-scalar ratio that measures the GW signal amplitude over the magnitude of scalar density fluctuations driving the formation of cosmic structures. Also, $n_T$ is the spectral index of tensor perturbations, and $\mathscr{k}_0= 0.05\ \text{Mpc}^{-1}$ is the pivot scale. The latter has been used in the most recent Planck-CMB analyses to place limits on $r$ \cite{Planck:2018vyg,Planck:2018jri}.

The parametric form given in  Eq.~\eqref{parametrization flat PS} takes into account deviations from the scale-invariant predictions of perfect de Sitter inflation, as they occur in the standard slow-roll scenario \cite{Liddle:2000cg}.
In fact, inflation is expected to end and, thus, spacetime has to deviate from the ideal de Sitter model that is characterized by eternal inflation.
The combination of Eqs.~\eqref{eq:curved PS} and \eqref{parametrization flat PS} yields a general parametrization of the primordial power spectrum that includes the effects of non-vanishing curvature:
\begin{equation}
\mathcal P_T(\mathscr{k})= r A_S \left(\frac{\mathscr{k}}{\mathscr{k}_0}\right)^{n_T} \frac{\mathscr{k}^4}{\mathscr{k}^4-\mathcal{K}^2}\,.
    \label{eq:param_PS}
\end{equation}

Furthermore, we consider the effective degrees of freedom of relativistic species in the primordial plasma, so that we can write the energy density and the entropy density as, respectively, \cite{Kolb:1990vq}
\begin{align}
    \rho_\text{rad}=\dfrac{\pi^2}{30}g_{\text{eff},\rho}T^4\,, \quad 
     s=\dfrac{2\pi^2}{45}g_{\text{eff},s} T^3\,,
\end{align}
where 
\begin{align}
    g_{\text{eff},\rho}&=\sum_{i=\rm bosons}g_i\left(\dfrac{T_i}{T}\right)^4+\dfrac{7}{8}\sum_{i=\rm fermions}g_i\left(\dfrac{T_i}{T}\right)^4\,, \\
     g_{\text{eff},s}&=\sum_{i=\rm bosons}g_i\left(\dfrac{T_i}{T}\right)^3+\dfrac{7}{8}\sum_{i=\rm fermions}g_i\left(\dfrac{T_i}{T}\right)^3\,.
\end{align}

The amplitude of GWs we observe today could be studied by analyzing the modes that entered the horizon in the RD epoch. Thus, the WKB approximation proves suitable for describing the transfer function after the modes reenter the horizon. Within such approximation, one has \cite{Saikawa:2018rcs}
\begin{equation}
    \mathscr{T}^{\prime}(\tau)^2\approx \dfrac{\mathcal{H}_\star^2\, a_\star^2}{2a(\tau)^2}\,,
\end{equation}
where $\mathcal{H}_\star\equiv\mathcal{H}(\tau_\star)=\mathscr{k}$, and $a_\star$ is the scale factor at the time of horizon crossing, $\tau_\star$. Therefore, Eq.~\eqref{eq: spectral energy density} yields
\begin{equation}
    \Omega_\text{GW}(\tau)=\dfrac{\mathcal{P}_T}{24}\left[\dfrac{\mathcal{H}_\star \, a_\star}{\mathcal{H}(\tau)a(\tau)}\right]^2\,.
    \label{eq: spectral energy density species}
\end{equation}
At the time of horizon crossing, the first Friedmann equation reads
\begin{equation}
    \left(\frac{\mathcal{H}_\star}{\mathcal{H}_0}\right)^2=\frac{\Omega_{\gamma,0}}{2a_\star^2}g_{\text{eff},\rho\star}\left(\frac{g_{\text{eff},s0}}{g_{\text{eff},s\star}}\right)^{4/3}+\Omega_{\mathcal{K},0}\,,
    \label{eq: H hc 2}
\end{equation}
with $g_{\text{eff},\rho\star}\equiv g_{\text{eff},\rho}(T_\star)$, where  $T_\star$ is the universe's temperature at horizon crossing, whereas $g_{\text{eff},s0}\equiv g_{\text{eff},s}(T_0)$, with $T_{0}$ being the current temperature of the CMB. Moreover, $\Omega_{\gamma,0}\equiv \frac{\pi^2}{45}\frac{T_0^4}{\mathcal{H}_0^2}$ and $\Omega_{\mathcal{K},0}\equiv -\frac{\mathcal{K}}{\mathcal{H}_0^2}$ are the present fraction densities of photons and curvature, respectively.
Hence, the relic energy density of primordial GWs is finally given by
\begin{equation}
    \Omega_{\text{GW},0}(\mathscr{k})=\mathscr{G}\, \mathcal P_T(\mathscr{k})\,,
\end{equation}
where we have defined
\begin{equation}
    \mathscr{G}\equiv \left(\frac{g_{\text{eff},s0}}{g_{\text{eff},s\star}}\right)^{\frac{2}{3}} \left[\frac{\Omega_{\gamma,0}}{48}g_{\text{eff},\rho\star}\left(\frac{g_{\text{eff},s0}}{g_{\text{eff},s\star}}\right)^\frac{2}{3} + \frac{\Omega_{\mathcal{K},0}}{24}\left(  \dfrac{T_{\gamma,0}}{T_\star}\right)^2\right].
\end{equation}

\subsection{Reheating}

According to the standard reheating scenario, after the inflationary epoch, the universe undergoes a phase characterized by oscillations of the inflaton field, followed by the RD epoch. 
The modifications in the spectral shape induced by the radiation-matter equivalence and the reheating phase are taken into account as \cite{Nakayama:2008wy}
    \begin{equation}
    \Omega_{\text{GW},0}=\mathscr{G}\,\mathcal{P}_T\,\mathscr{T}_\text{int}^2\,\mathscr{T}_\text{rh}^2\,,
    \label{eq:spectrum}
\end{equation}
where $\mathscr{T}_\text{int}$ is the transfer function in the intermediate regime between the RD and MD epochs, for different spatial geometries, as given in Ref.~\cite{DAgostino:2023tgm}. 
Moreover, $\mathscr{T}_\text{rh}$ is the transfer function in the reheating phase that is well approximated by the following fitting formula
\cite{Kuroyanagi:2020sfw,NANOGrav:2023hvm}:
\begin{equation}
    \mathscr{T}_\text{rh}^2\simeq \Theta(\mathscr{k}-\mathscr{k}_\text{end}) \left(1-0.22\,x_\text{rh}^{3/2}+0.65\,x_\text{rh}^2\right)^{-1}\,.
\end{equation}
Here, the Heaviside function $\Theta$ is introduced to specify the GWB spectrum endpoint, namely at $\mathscr k=\mathscr k_\text{end}$, when inflation ends and reheating takes place:
\begin{equation}
\mathscr{k}_\text{end}=\left[\frac{\pi^2 g_\text{eff}^\text{(rh)}}{90}\right]^{1/2} \left[\dfrac{g_{\text{eff},s0}}{g^\text{(rh)}_{\text{eff},s}}\right]^{1/3} T_0\, (T_\text{rh} H_\text{end})^{1/3}\,,
\end{equation}
where $H_\text{end}$ is the Hubble rate at the end of inflation, while $g_{\text{eff}}^\text{(rh)}\equiv g_{\text{eff}}(T_\text{rh})$, $g_{\text{eff},s}^\text{(rh)}\equiv g_{\text{eff},s}(T_\text{rh})$ and $T_\text{rh}$ is the reheating temperature, right before the universe enters the RD epoch. 
Additionally, $x_\text{rh} \equiv \mathscr k/\mathscr k_\text{rh}$, with $\mathscr k_\text{rh}$ being the typical wavenumber at the end of reheating:
\begin{equation}
    \mathscr k_\text{rh}=\left[\frac{\pi^2 g_\text{eff}^\text{(rh)}}{90}\right]^{1/2} \left[\dfrac{g_{\text{eff},s0}}{g^\text{(rh)}_{\text{eff},s}}\right]^{1/3}  T_\text{rh} T_0\,.
    \label{eq:k_rh}
\end{equation}

\section{Methodology}
\label{sec:methodology}

The NANOGrav 15-year dataset includes the pulse time of arrivals (TOAs) of 68-millisecond pulsars. With a timing baseline of 3 years, 67 of these pulsars remain viable for processing \cite{NANOGrav:2023hde,NANOGrav:2023hvm}.
Specifically, in the present analysis, we use the pulsar timing residuals ($\bm{\delta t}$) to acquire information from the primordial power spectrum.
The timing residuals represent the discrepancy between the observed TOAs and those predicted by the pulsar timing model. In particular, as described in Ref.~\cite{NANOGrav:2023hvm}, the timing residuals can be modeled as 
\begin{equation}
\label{eq: timing residual}
   \bm{\delta t = n  + M \epsilon  + F a} \,.
\end{equation}
Here, {$\bm n$} represents the contribution of the white noise, which is assumed to be a normal random variable with zero mean. The covariance matrix of the white noise, for a given receiver/back-end combination $I$, is
\begin{equation}
    \langle n_i n_j\rangle = \mathcal{F}_{I}^2\left[\sigma^2_{i,S/N} +\mathcal{Q}^2_I\right] + \mathcal{J}_I^2\mathcal{U}_{ij}\, ,
\end{equation}
where $i$ and $j$ label the TOAs and  $\sigma^2_{i,S/N}$ is the TOA uncertainty relative to the $i$-th observation. Moreover, $\mathcal{F}_{I}$, $\mathcal{Q}_I$ and $\mathcal{J}_I$ are the extra factor, quadrature and  correlation parameters, respectively, while $\mathcal{U}$ is a block-diagonal matrix with unitarity values for TOAs that belong to the same observing time, and null values for all the other elements \cite{NANOGrav:2023hde}.

The $\bm{M \epsilon}$ term in Eq.~\eqref{eq: timing residual} measures the departures from the initial best-fit values of the $m$ timing-ephemeris parameters \cite{NANOGrav:2020tig}. Specifically, $\bm{M}$ is a $m\times N_\text{TOA} $ matrix including the partial derivatives of the TOAs over each $m$ parameter, calculated at the best-fit value, whereas the vector $\bm \epsilon$ contains the offsets to the best-fit values.

Finally, the $\bm{Fa}$ term is a combination of the pulsar-intrinsic red noise and the stochastic GWB signal. In particular, $\bm{F}$ is the design matrix accounting for the Fourier basis of frequencies $i/T_\text{obs}$, where $i$ indexes the harmonics of the basis and $T_\text{obs}$ is the timeline baseline.
In our analysis, we use $30$ frequencies to model the pulsar-intrinsic red noise and $14$ frequencies for the GWB. Indeed, observations show that the evidence for a GWB comes from the first $14$ frequency bins \cite{NANOGrav:2023gor}.
Moreover, the vector ${\bm a}$ includes the coefficients of the Fourier expansion that are taken as normally distributed random variables with zero mean and covariance matrix with coefficients
\begin{equation}
\label{eq : covariance}
    [\phi]_{(ai)(bj)} = \delta_{ij} \left(\Gamma_{ab}\Phi_i + \delta_{ab}\varphi_{a,i}\right),
\end{equation}
such that $\langle\bm{aa^\text{T}}\rangle=\bm \phi$.
Here, $a$ and $b$ label the pulsars, $i$ and $j$ index the frequency harmonics, while $\Gamma_{ab}$ measures correlations between pulsars $a$ and $b$, as a function of their sky angular separation \cite{Hellings:1983fr}.
Additionally, the term $\Phi_i$ parametrizes the contribution to the timing residual of the given GWB model. 
In particular, we focus on the GWB originating from an astrophysical source, such as an SMBHB population, and from a cosmological source, such as primordial GWs induced by inflation.
The GW spectrum from SMBHB is studied and tested in \cite{NANOGrav:2023smbhb}, where tensions between the NANOGrav dataset and the prediction of SMBHB models arise. Hence, we can test models that describe the GW spectrum generated during the inflation to fit the data better than the conventional SMBHB signal.
Finally, the coefficients $\varphi_a$ in Eq.~\eqref{eq : covariance} describe the pulsar-intrinsic red noise as
\begin{equation}
\label{eq: phi_a}
    \varphi_a (f) = \frac{A_a^2}{12 \pi^2} \frac{1}{T_\text{obs}}\left( \frac{f}{1\,\text{yr}^{-1}}\right)^{-\gamma_a}\text{yr}^3\,,
\end{equation}
such that $\varphi_{a,i}\equiv \varphi_a(i/T_\text{obs})$, with $i$ running over all frequencies. Here, $A_a$ and $\gamma_{a}$ are the red noise amplitude and spectral index, respectively, and the frequency $f$ is related to the wavenumber through the relation $f= \mathscr{k}/2\pi\,.$

Thus, we marginalize over all possible noise realization, namely over all possible values of $\bm a$ and $\bm \epsilon$. Doing so, the marginalized likelihood will depend only on the red noise parameter set, $\bm \theta$, and the model-dependent parameters encoded in $\Phi_i$. 
Therefore, the likelihood function reads
\begin{equation}
    \label{eq:likelihood}
    \mathcal{L}(\bm{\delta t}|\bm{\theta})=\frac{\exp{\left[-\frac{1}{2}\bm{\delta t}^\text{T}\bm{C}^{-1}\bm{\delta t}\right]}}{\sqrt{|2\pi\bm{C}}|}\,.
\end{equation}
Here, ${\bm C = \bm N +\bm{TBT}^\text{T}}$, where $\bm{N}$ is the white noise covariance matrix, and $\bm{T}=[\bm M,\bm F]$ is a block matrix.
Moreover, $\bm B=\text{diag}(\bm{\infty}, \bm{\phi})$, with $\bm{\infty}$ being the diagonal infinity matrix that is related to the flat prior assumption on the $\bm \epsilon$ parameters.
To speed up the calculations, as pointed out in Ref.~\cite{Lamb:2023jls}, we can fit directly our GWB model to the free spectrum of the PTA data. In particular, the free spectrum is given by the posterior distributions on $\Phi_i$ at each sampling frequency, $p(\Phi_i|\bm{\delta t})$. Hence, Eq.~\eqref{eq:likelihood} becomes
\begin{equation}
    \label{eq:likelihhod_ceffyl}
    \mathcal{L}(\bm{\delta t}|\bm\theta)=\prod_{i=1}^{14} \left.\frac{p(\Phi_i|\bm{\delta t})}{p(\Phi_i)}\right|_{\Phi_i = \Phi_{\text{GWB}}\left(i/T_\text{obs},\bm{\theta}\right)}\,,
\end{equation}
where $p(\Phi_i)$ is the prior probability for $\Phi_i$, while $\Phi_\text{GWB}\left(i/T_  \text{obs},\bm{\theta}\right)$ is the GWB spectrum depending on the model parameters.

\section{Results and discussion}
\label{sec:results}

We perform a Bayesian analysis by means of the Python package \texttt{PTArcade} \cite{Mitridate:2023oar}, which integrates new physics into the PTA data analysis package \texttt{ceffyl} \cite{Lamb:2023jls}. Specifically, \texttt{PTArcade} samples the posterior distribution through the Markov chain Monte Carlo (MCMC) algorithm implemented in the \texttt{PTMCMCSampler} package \cite{justin_ellis_2017_1037579}.

In our numerical procedure, we shall keep the $\Omega_{\mathcal{K},0}$ parameter independent from the Hubble constant, which we fix to the latest estimate of the Planck collaboration \cite{Planck:2018vyg}, $h=0.67$, where $h\equiv \mathcal{H}_0$/(100 km s$^{-1}$ Mpc$^{-1})$.
We thus label as $\mathcal{K}$-GW the spatially curved GWB described by Eq.~\eqref{eq:spectrum}.
The parameters $r$ and $n_T$ are treated as independent variables throughout the numerical sampling.
In particular, we set uniform priors both on $\log_{10}(r)$, i.e., $\mathcal{U}(-40,0)$, and on $n_T$, i.e., $\mathcal{U}(0,6)$.
Furthermore, from Eq.~\eqref{eq:k_rh}, we can estimate the frequency at reheating phase: $f_{\text{rh}}\sim 30\ \text{nHz}\ (T_{\text{rh}}/1\,\text{GeV})$. Hence, we impose a uniform prior on $\log_{10}(T_{\text{rh}}/1\,\text{GeV}) $, namely $\mathcal{U}(-3,3)$. Finally, we sample $\Omega_{\mathcal{K},0}$ uniformly in the range $[-1,1$].
\begin{figure}
    \centering
    \includegraphics[width=0.48\textwidth]{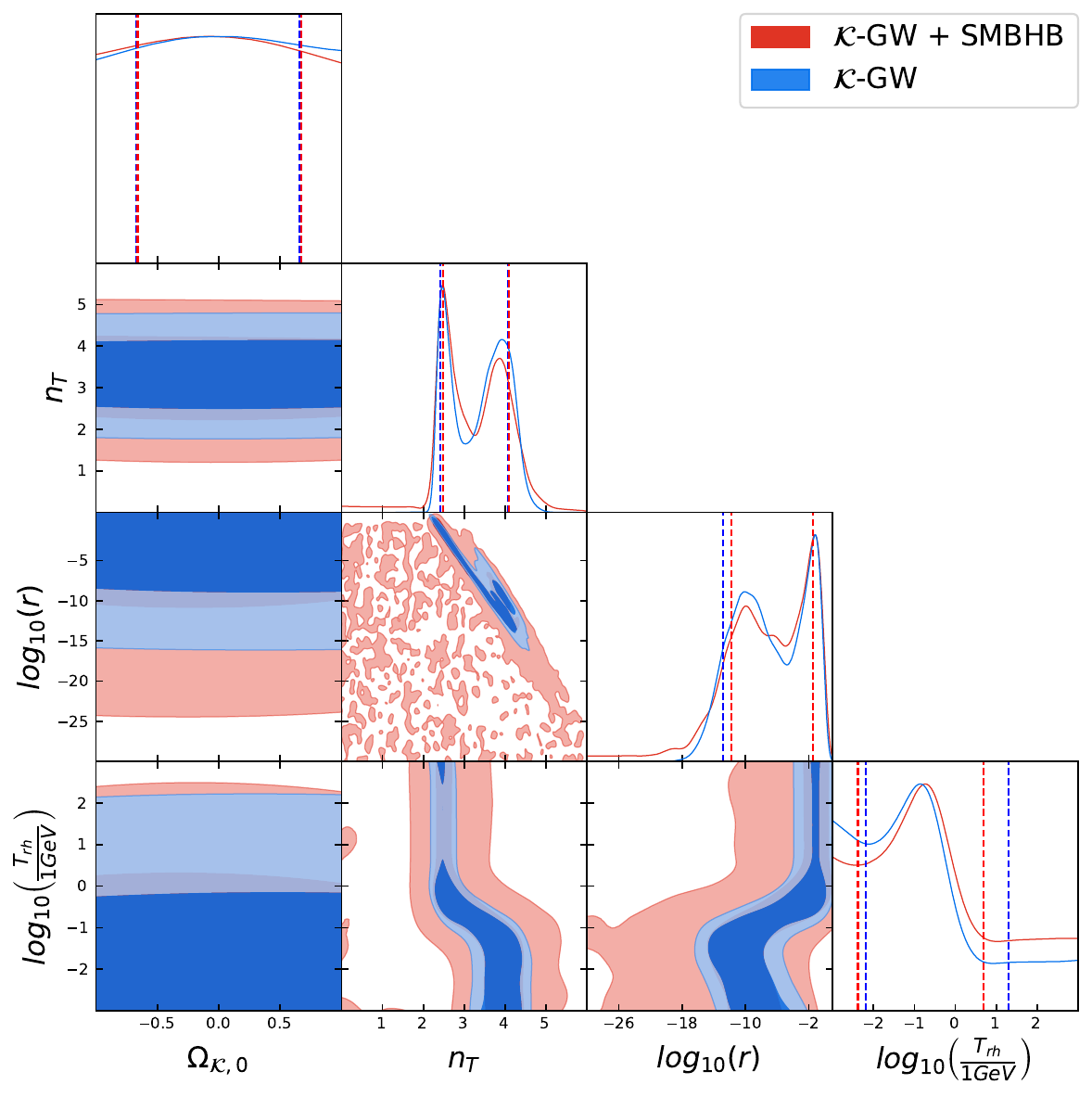}
    \caption{Marginalized 68\% and 95\% CL contours, and posterior distributions, for the free parameters of the $\mathcal{K}$-GW  and $\mathcal{K}$-GW\,+\,SMBHB  models. The dashed lines enclose the $1\sigma$ regions.}
    \label{fig:corner KGw + bg}
\end{figure}

In Fig.~\ref{fig:corner KGw + bg}, we show the $1\sigma$-$2\sigma$ contour plots and the posterior distributions for the $\mathcal{K}$-GW spectrum and the combined spectrum originating from inflation plus the SMBHB signal.
We can see that the behaviors of $\log_{10}(r)$, $n_T$ and $\log_{10}(T_{\text{rh}}/1\,\text{GeV})$ are analogous to those emerged from the analysis of NANOGrav \cite{NANOGrav:2023hvm}.
Specifically, we note a strong covariance between the $n_T$ and $\log_{10}(r)$, and a bimodal distribution for both the marginalized posteriors. These features remain unchanged for the $\mathcal{K}$-GW  and $\mathcal{K}$-GW\,+\,SMBHB spectra, respectively. 
In the 2D contour plots of the pairs $\{\log_{10}(T_{\text{rh}}/1\,\text{GeV}),n_T\}$ and $\{\log_{10}(T_{\text{rh}}/1\,\text{GeV}),\log_{10}(r)\}$, the bimodality induces a reflection symmetry with respect to the points $(-0.6, 3.25)$ and $(-0.6, -6)$ in the $\{\log_{10}(T_{\text{rh}}/1\,\text{GeV}),n_T\}$ and $\{\log_{10}(T_{\text{rh}}/1\,\text{GeV}),\log_{10}(r)\}$ planes, respectively.
In analogy with the analysis made in Ref.~\cite{NANOGrav:2023hde}, we highlight two regimes: $T \ll  1$ GeV and $T\gg 1$ GeV.
In the first one, the reheating frequency $f_\text{rh}$ is below the PTA frequencies and the GW spectrum in the observed band is composed of tensor modes that re-entered the horizon during reheating after inflation.
The second regime is characterized by a $f_\text{rh}$  greater than the frequencies in the PTA band. In this case, the GW spectrum comprises tensor modes that re-entered the horizon during the radiation era. 
Our results indicate that the spatial curvature parameter is constrained to $\Omega_{\mathcal{K},0}= 0.00 \pm 0.67$ at the 68\% confidence level (CL).

\begin{figure}
    \centering
    \includegraphics[width=0.48\textwidth]{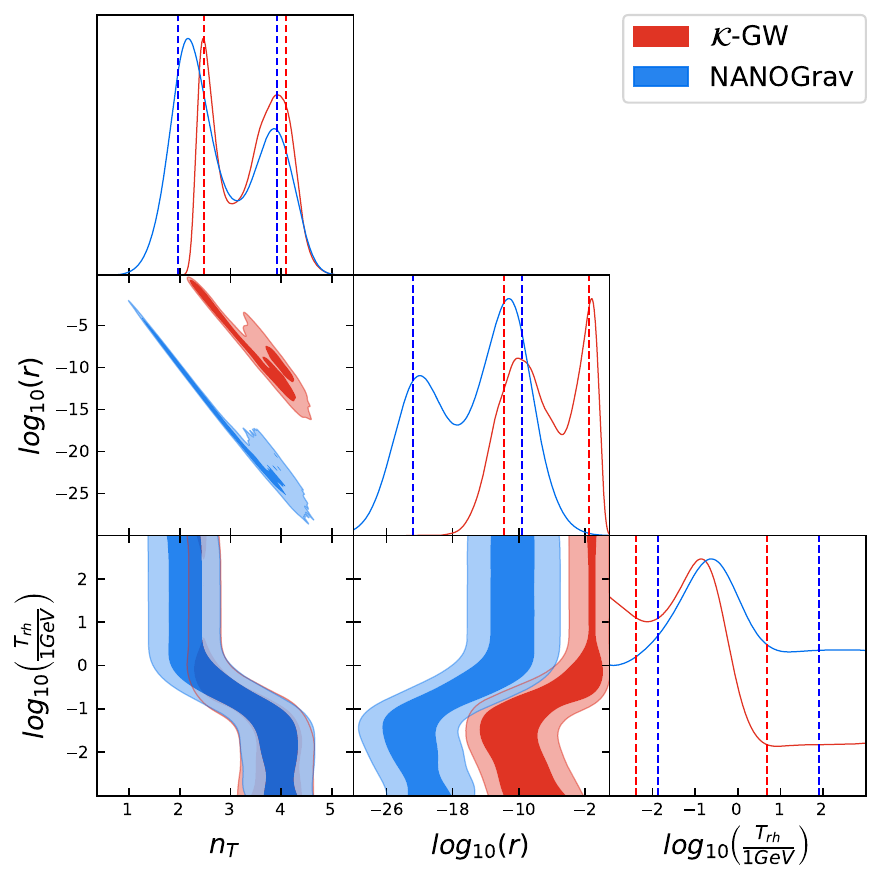}
    \caption{Comparison between the 68\% and 95\% CL results of our analysis and those of the NANOGrav collaboration \cite{NANOGrav:2023hde}.}
    \label{fig: comparison with Nanograv}
\end{figure}

In Fig.~\ref{fig: comparison with Nanograv}, we compare the results on the common parameters of the $\mathcal{K}$-GW scenario and the cosmic inflation spectrum considered by NANOGrav, i.e., $\log_{10}(r)$, $n_T$ and $\log_{10}(T_{\text{rh}}/1\,\text{GeV})$.
A common feature of both models is the bimodal distributions for $\log_{10}(r)$ and $n_T$, leading to two different regimes for $T_{\text{rh}}$. 
Nevertheless, we notice that the posterior distribution of $\log_{10}(r)$ in our model is shifted with respect to the NANOGrav findings, resulting in a factor of $\sim 10^{11}$ discrepancy for the most likely values of $r$.
However, both estimates agree with the Planck constraint, namely $r< 0.06$ \cite{Planck:2018jri}.
Moreover, the NANOGrav collaboration constrains the value of the strain amplitude to $2.4^{+0.7}_{-0.6} \times 10^{-15}$  at a reference frequency of 1 yr$^{-1}$ \cite{NANOGrav:2023gor}. This bound induces a limit on the spectral index $n_T$. Specifically, when $T \ll  1$ GeV we recover the peak at $n_T = 4 $ in the posterior distribution, in analogy with the NANOGrav analysis \cite{NANOGrav:2023hvm}. On the other hand, when $T \gg  1$ GeV, we find a peak at $n_T = 2.5 $, while the NANOGrav posterior shows a peak at $n_T =2$ \cite{NANOGrav:2023hvm}.

\begin{figure}
    \centering
    \includegraphics[width=0.48\textwidth]{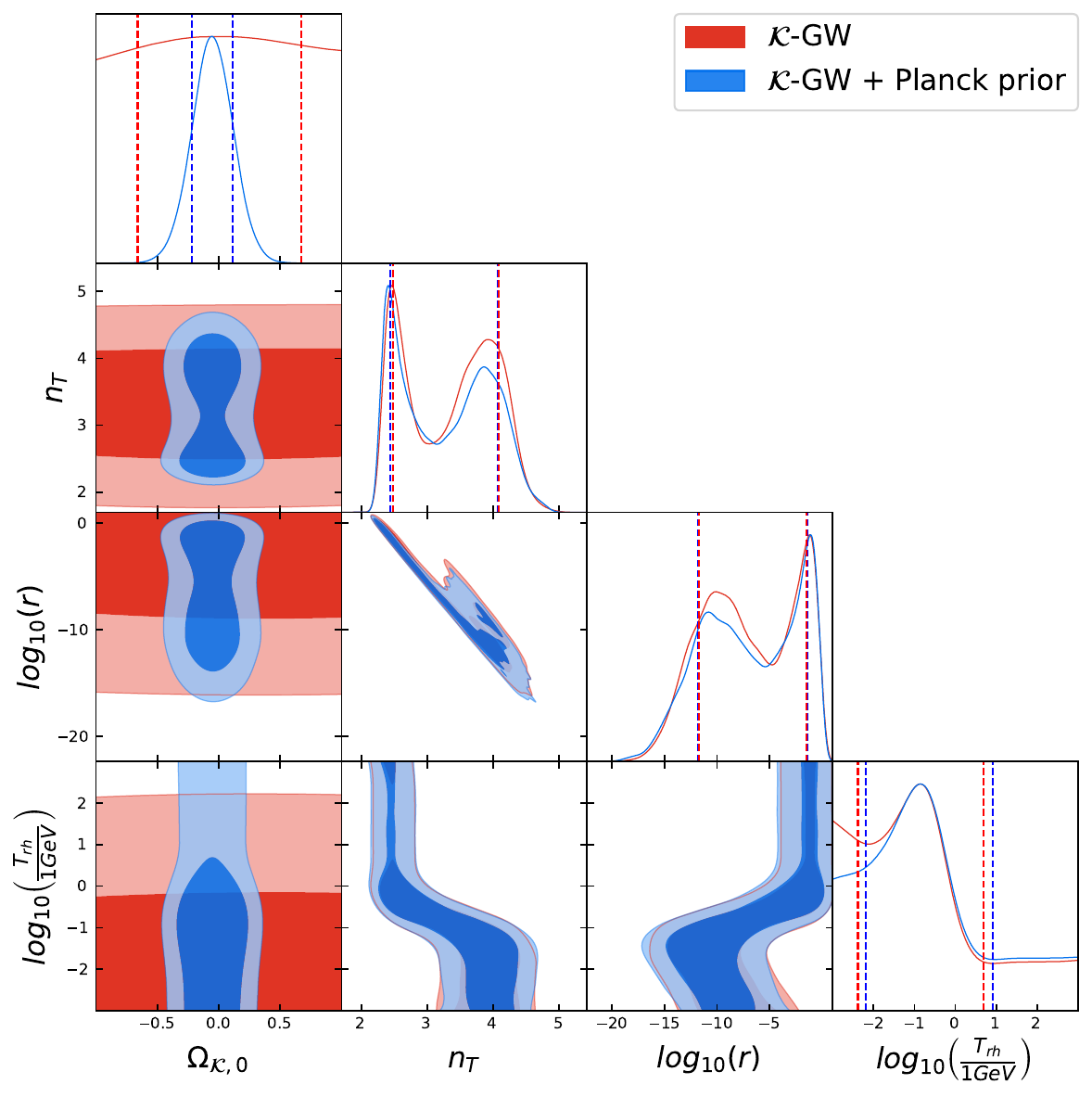}
    \caption{Comparison between the 68\% and 95\% CL results obtained by assuming a uniform (red) and a Gaussian (blue) prior on $\Omega_{\mathcal{K},0}$. The latter is based on the constraint given by the Planck collaboration \cite{Planck:2018jri}. }
    \label{fig: planck prior}
\end{figure}

Furthermore, we analyze the case when a Gaussian prior is imposed on $\Omega_{\mathcal{K},0}$. In particular, for the mean and the variance of the Gaussian distribution, we consider the best-fit and the 99\% CL values obtained by Planck, i.e., $-0.06$ and $0.18$, respectively \cite{Planck:2018jri}.
In Fig.~\ref{fig: planck prior}, we thus compare the results for the $\mathcal{K}$-GW spectrum in the case of uniform and Gaussian priors on $\Omega_{\mathcal{K},0}$.
We note that the values of $\log_{10}(r)$, $n_T$ and $\log_{10}(T_{\text{rh}}/1\,\text{GeV})$ are quite independent from the prior on  $\Omega_{\mathcal{K},0}$, as the corresponding 1D posterior distributions and the 2D contours overlap. On the other hand, such analysis allows us to improve the accuracy on $\Omega_{\mathcal{K},0}$ by a factor $\sim 4$: $\Omega_{\mathcal{K},0}= -0.05 \pm 0.17$ (68\% CL).
The latter shows the significant role played by the priors in the present analysis.

\subsection{Consistency checks}

Here, we conduct two consistency checks to validate our study and ensure the absence of possible numerical artifacts in our analysis.  

\begin{figure}
    \includegraphics[width=0.48\textwidth]{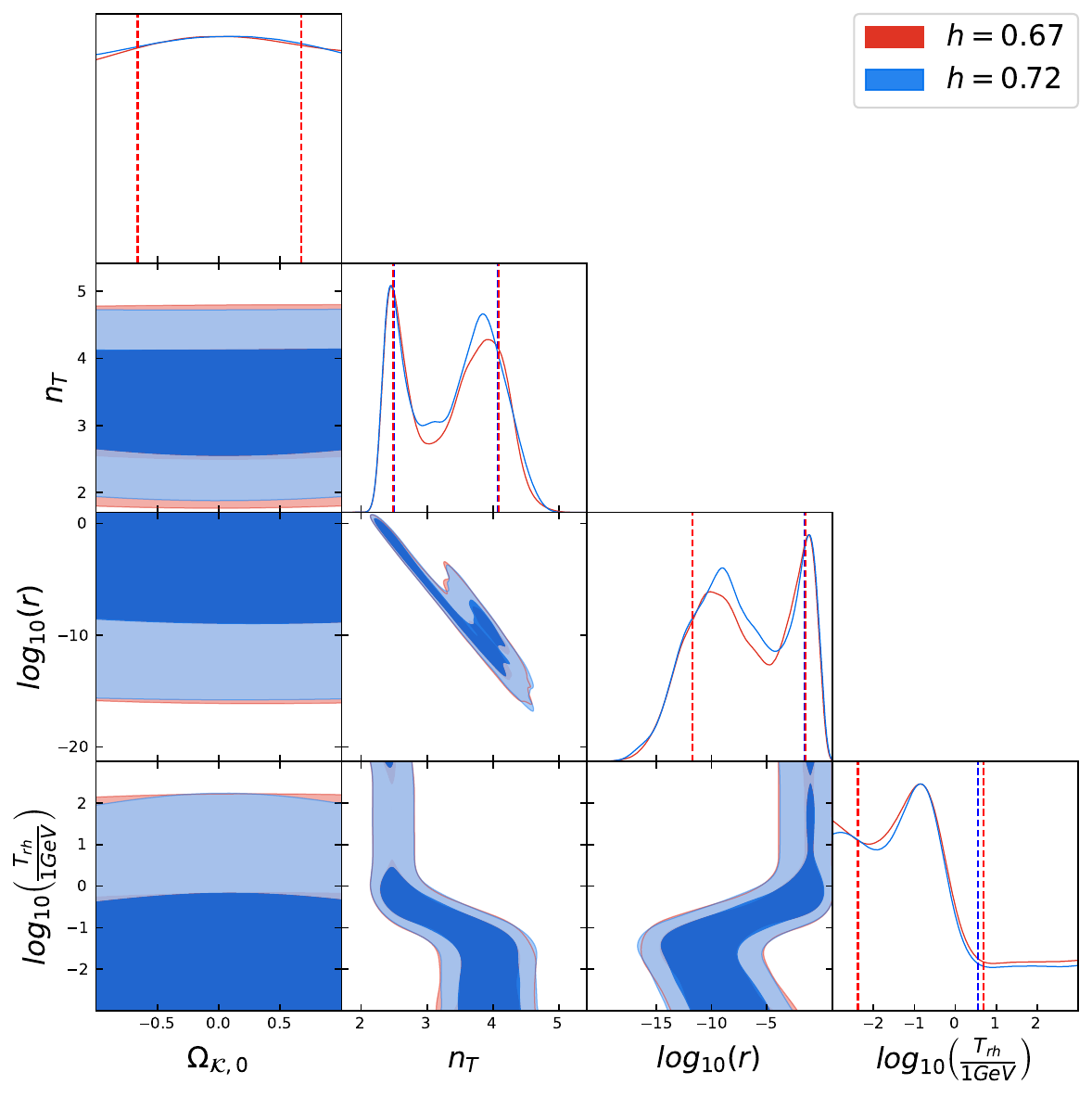} 
    \caption{Marginalized 68\% and 95\% CL contours, and posterior distributions, for the free parameters of the $\mathcal{K}$-GW model under different assumptions for $h$.} 
    \label{fig: corner h}
\end{figure}

We first examine the impact of our $h$ assumption. In the main analysis, we set $h=0.67$, in agreement with the Planck result \cite{Planck:2018jri}. On the other hand, one may consider the direct measurement of the Hubble constant based on local Cepheids \cite{Riess:2021jrx}. In this case, we can assume $h=0.73$ and
compare the posterior distributions obtained with the two different values of the Hubble constant (see Fig.~\ref{fig: corner h}). We notice that the choice of $\mathcal{H}_0$ does not actually affect our results. 

\begin{figure}
    \includegraphics[width=0.48\textwidth]{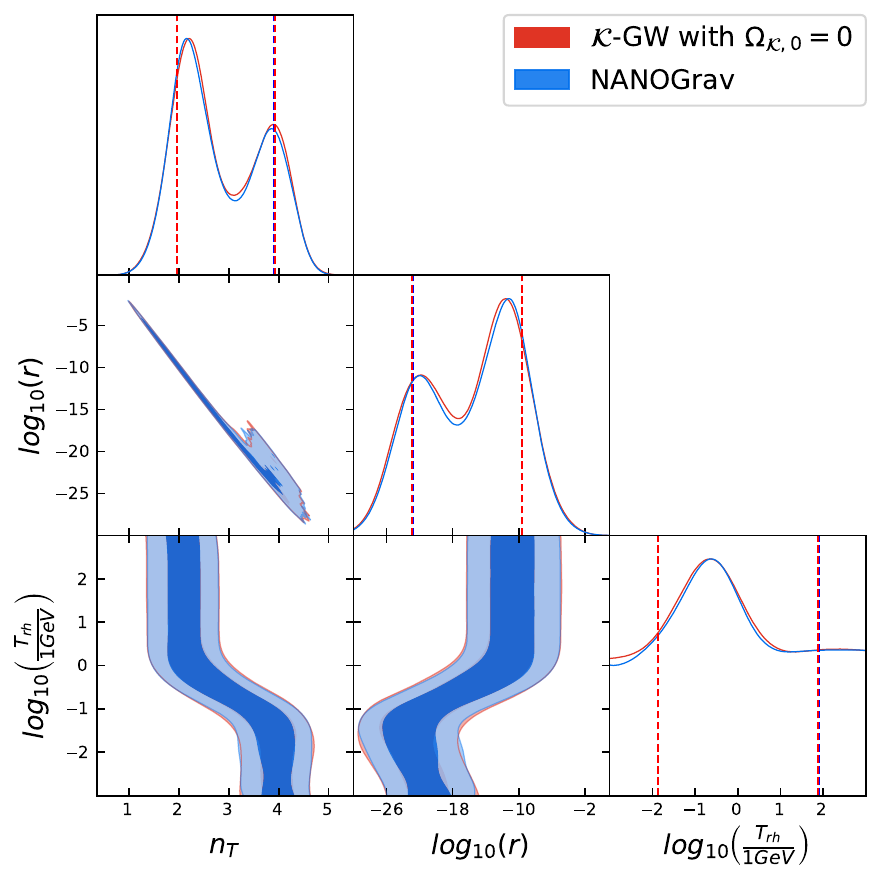}
    \caption{Comparison between the 68\% and 95\% CL predictions of the $\mathcal{K}$-GW model with $\Omega_{\mathcal{K},0}=0$ and the results of the NANOGrav collaboration \cite{NANOGrav:2023hde}.}
    \label{fig: numerical check}    
\end{figure}

The second check consists of fixing $\Omega_{\mathcal{K},0}=0$ in the $\mathcal{K}$-GW model. As shown in Fig.~\ref{fig: numerical check}, the contours and the posterior distributions fully overlap with the NANOGrav results that were obtained in the case of vanishing spatial curvature. Once again, this confirms the goodness of our analysis and, therefore, validates the new findings related to the non-flat universe scenario.

\section{Final remarks}
\label{sec:conclusions}

In this study, we investigated primordial GWs that originated during inflation and their propagation in the subsequent radiation and matter epochs. Specifically, we considered the solutions to the Friedmann equations for an isotropic and homogeneous background universe with non-vanishing spatial curvature. Then, we introduced linear tensor perturbations around the FLRW metric to account for the GW evolution under different universe's geometry. With the help of transfer functions and suitable initial conditions, we obtained the energy density and the power spectrum of primordial GWs. 

Therefore, we proposed a new parametrization of the tensor power spectrum that includes a correction factor due to the presence of non-zero spatial curvature, in addition to the tensor-to-scalar ratio and the spectral index typical of the standard flat scenario. We then expressed the relic GW energy density in terms of effective degrees of freedom contributing to the entropy and density of relativistic particles. Moreover, we considered the modifications in the spectral shape induced by the radiation-matter equivalence and the reheating phase.

Given the above theoretical scenario, we revisited the constraints on spatial curvature in light of the recent observations of a stochastic GWB. For this purpose, we employed the NANOGrav 15-year dataset release and performed a Bayesian analysis of the newly proposed parametrization of the tensor power spectrum. In particular, assuming uniform flat priors on the free parameters of the models, we found $\Omega_{\mathcal{K},0}= 0.00 \pm 0.67$ (68\% CL).  
Therefore, to effectively constrain the geometry of the universe, we assumed a $3\sigma$ Planck prior on $\Omega_{\mathcal{K},0}$. In doing so, we obtained $\Omega_{\mathcal{K},0}= -0.05 \pm 0.17$ (68\% CL).
Furthermore, we found bimodal distributions for both $\log_{10}r$ and $n_T$, whose behaviors are analogous to those obtained by the NANOGrav collaboration.
However, our posterior on $\log_{10}r$ is shifted with respect to the NANOGrav results by several orders of magnitude.
Nevertheless, our results on $r$ are in agreement with the Planck-CMB limits.

Finally, we carried out some consistency checks to validate our results. First, we investigated the impact of the Hubble constant value on the final numerical outcomes. Specifically, we showed that the same conclusions can be achieved by either assuming the Planck value or the local estimate for $\mathcal{H}_0$. Then, we examined potential numerical artifacts by analyzing our GWB model under the assumption of vanishing curvature. In doing so, we replicated the NANOGrav results, thereby confirming the validity of the new findings associated with the non-flat universe scenario.

In the context of recent debates on possible evidence of non-zero curvature, the present study provides further support for a flat universe, in agreement with the standard predictions of the CMB anisotropies.
In conclusion, it is important to note that our findings are derived from a simplified power-law form of the tensor power spectrum. Exploring more sophisticated cases would be interesting to validate the current results. Additionally, while at present the NANOGrav observations alone seem to have a marginal impact on constraining spatial curvature, future releases of PTA data in the coming years may enhance the accuracy of all background parameters.

\acknowledgments

The authors want to express their gratitude to the anonymous referee for her/his valuable comments and suggestions that significantly helped to improve the quality of the manuscript.
The authors acknowledge the financial support of the Istituto Nazionale di Fisica Nucleare  (INFN) - Sezione di Napoli, {\it iniziative specifiche} QGSKY, MOONLIGHT and TEONGRAV. 
R.D. acknowledges work from COST Action CA21136 - Addressing observational tensions in cosmology with systematics and fundamental physics (CosmoVerse).
D.V. acknowledges the FCT Project No. PTDC/FIS-AST/0054/2021.


\bibliography{references}

\end{document}